# ACO-ESSVHOA - Ant Colony Optimization based Multi-Criteria Decision Making for Efficient Signal Selection in Mobile Vertical Handoff

A. BHUVANESWARI,
Asst. Professor,
Dept. of Computer Science,
Cauvery College for Women,
Trichy, Tamil Nadu, India.
prkrizbhu@yahoo.co.in

E. George Dharma Prakash Raj
Asst. Professor,
Dept. of Computer Science,
School of Computer Science and Engineering,
Bharathidasan University, Trichy, Tamil Nadu, India.

V. SINTHU JANITA PRAKASH
Asst. Professor,
Dept. of Computer Science,
Cauvery College for Women,
Trichy, Tamil Nadu, India.

*Abstract* -The process of Vertical handoff has become one of the major components of today's wireless environment due to the availability of the vast variety of signals. The decision for a handoff should be performed catering to the needs of the current transmission that is being carried out. Our paper describes a modified Ant Colony Optimization (ACO) based handoff mechanism, that considers multiple criteria in its decision making process rather than a single parameter (pheromone intensity). In general, ACO considers the pheromone intensity and the evaporation rates as the parameters for selecting a route. In this paper, we describe a mechanism that determines the evaporation rates of each path connected to the source using various criteria, which in turn reflects on the pheromone levels present in the path and hence the probability of selecting that route. Experiments show that our process exhibits better convergence rates, hence better usability.

*Keywords: Multi Criteria Decision Making; Ant Colony Optimization; Vertical Handoff; Evolutionary Computation*

I. INTRODUCTION

Due to the fact that all wireless devices in motion can have reduced signal rates at some point, handoff has become an essential procedure. The basic component of any handoff mechanism is the handoff decision making phase that determines whether the call is to be handed off or not, and if yes, to which signal provider the handoff should take place such that the provider meets all the requirements of the current transmission, and will have a high probability of handling the call for large periods of time without the need for a handoff.

The decision making process in handoff mechanism can in general be described as a scheduling mechanism that requires optimal resources (here antenna/base signal). The optimality in this case is determined by determining the requirements of the user. The signal sources that match these requirements are considered for utilization. Hence the handoff decision making process is considered a special case of scheduling algorithms.

Usage of statistical scheduling algorithms for the class of optimization algorithms is in general discouraged due to the fact that the time taken by these algorithms cannot be determined. When the input is very large, the processing times of these algorithms becomes very high and hence are not reliable. Moreover the scheduling class of algorithms has the property of accepting a near optimal solution, rather than a point accurate solution. The accuracy acceptance rate of these algorithms are moderate, hence a heuristic based algorithm would perform well in providing a solution at the stipulated time.

The remainder of this paper is structured as follows; section II provides the related works, section III provides an overview of the process, section IV presents the actual process, section V provides the results and VI concludes the study.

II. RELATED WORKS

Evolutionary algorithms[1] are the class of algorithms that evolved from nature. These algorithms have been designed inspired by certain elements from nature, such as ants, bees, etc. In general, these classes of algorithms are used for a single object decision making[2]. When observed closely, all real time problems are multi objective in nature. i.e. the result must satisfy multiple constraints and not just a single constraint. The evolution based programming is performed in two steps. Initially a population set is selected, from which the next generation is derived using the below formula

$$x(t+1) = s\big(v(x(t))\big) \qquad (1)$$

where, *x [t+1]* denotes the new population, *t* denotes time, *s* denotes the selection set, *v* denotes the random variation coefficient and *x(t)* denotes the population at time *t*. Evolutionary algorithms use computer based problem solving techniques. Eg. Genetic Algorithms, Evolutionary programming, Genetic Programming, etc. These algorithms in general, tend to generate a Pareto Optimal Set containing the solutions. Since we do not obtain a single solution using these approaches, a Pareto Optimal Set is used.

A heuristic based Ant Colony Optimization Algorithm is described in [3]. It uses the principle of Marco Dorigo, proposed in 1992 [4]. It modifies the algorithm described in [4] to incorporate multiple properties as conditional considerations to find the optimal route for the travelling salesman problem. [5] Describes a fuzzy based multi criteria decision making process. It describes the importance of Fuzzy Logic in the MCDM process. It describes the problems to be faced while working on multiple criteria, techniques to be used for solving these problems and the use of fuzzy logic in programming evolutionary systems. It describes the process of Fuzzy Multiple Criteria Decision Making and problem solving using Artificial Neural Networks, Genetic and Evolutionary techniques.

An Ant Colony workflow scheduling is described in [6]. Workflow is a sequence of tasks required to complete a single process. This in its general case is mapped to the optimization problem of ACO. [6] uses a time varying topology as its base and uses ACO to provide a process execution sequence that best helps in providing an efficient workflow.

An improved Ant Colony Algorithm that helps in scheduling of independent tasks is proposed in [7]. It uses the Min-Min algorithm to initialize the pheromone values, instead of using random pheromone values. It provides mechanisms that slow down the solving speed. Slowing down the process provides a reduced probability of premature pheromone conversion. [8] Provides a fuzzy reputation based Ant Colony algorithm. It uses the node reputation as a parameter for determining the optimal solution. These reputations or trust values are considered in the form of fuzzy numerical for easy computation.

[9] presents an adaptive Ant Colony algorithm for scheduling. A measure of adaptiveness is included by modifying the evaporation rate of the pheromone deposit instead of a single fixed value. The initial evaporation rate is set to a very minimal value, to make sure that it does not drop down to zero. The evaporation rate is also controlled such that very high or very low pheromone evaporations are avoided. A moderate evaporation parameter is maintained throughout the process. This helps in slow convergence of the system. Evolution slowdown is performed such that convergence is made stable. Hence it helps in the process converging to an optimal solution. Further, a positive feedback is maintained such that the evolution is gravitating towards the optimal solution. The problem with this approach is that it can sometimes lead to stagnation.

A chaos based ant colony optimization method is described in [10]. It uses a combination of ACO and chaos theory. It describes the system as a non linear discrete chaotic system. It provides mechanisms to synchronize two chaotic systems. This methodology is used for continuous domains. It uses a static task scheduling model. It does not facilitate communication and does not use priorities. Chaotic initialization of data and chaotic perturbation is performed. It finds solutions to each task and find the sequence with the biggest value of the objective function and updates it.

A multi agent based distributed ACO is described in [11]. It provides a modified ACO algorithm that can be sued in a distributed environment. Every ant is designing its own processor, and agents find the shortest path for resource discovery. This is performed on the basis of cost. Tour cost sharing will result in optimal solutions. This is a sequential model, hence time consumption is considerably higher.

A Parallelized ACO is proposed in [12], in which every group is assigned a parallel processor. Local best is communicated periodically to the partner. Partner determination is performed dynamically; hence the process yields different results at various points of time. Inefficient time intervals have the probability of leading to unnecessary data transfer. Partner determination leads to loss of data from other nodes. If the same partner is determined overtime, this will lead to unnecessary bottleneck. [13] describes a QoS aware Vertical Handoff decision making mechanism that uses Artificial Neural Networks for performing the decision process. [14] describes a statistical analysis of MCDM, the AHP. This technique is used for performing the process of multi criteria decision making using the user defined properties. This statistical method helps in determining the best solution, which can be used as a benchmark for analyzing the heuristic based algorithms. The paper [16] describes the AHP based channel selection based on QoS requirements of different traffic type among the WiFi, WiMAX and CDMA channels.

III. PROCESS- AN OVERVIEW

The described methodology uses Ant Colony Optimization as the base for finding the optimal channel. Since ACO in general is a single criteria decision making methodology, since our problem at hand requires an optimization solution taking into account multiple criteria (properties of a network), we use a modified form of the ACO with multi-criteria decision making process. The process of optimization begins when the call connection is initiated. The base node, the destinations and the ants are initialized, and the optimization process is carried out until call disconnection. WiFi, WiMAX and CDMA are considered as the base signals. The various signal properties discussed in [15] are used for performing the optimization process. By the implicit nature of the ACO, we can guarantee optimized handoff.

IV. ACO-ESSVHO

The process of monitoring begins when the initial network connection is triggered. The usual mechanism of connection happens when a device requires the services of a wireless network. The services requested by different users may vary, hence the required connection limitations also vary with respect to the users. The transmission of throughput sensitive data requires a large bandwidth, but is not time limited, while the requirements of delay sensitive or real time traffic[17] requires faster transmission and the bandwidth requirements are compromised. Hence, even during the initial call connection, the user requires a signal that fits their needs. Here, the base signals considered for processing include CDMA, WiFi and WiMAX. The quality of services provided by each of them varies considerably, hence these are considered. In reality, there may be many such antennae providing similar capabilities. In our current process, we maintain a stable set of three signals. Since the base properties of all the antennae providing this signal would be the same, the actual physical antenna selection is performed after the optimization process. Hence our algorithm provides flexibility to incorporate any number of signal sources, which come under the same base category.

After the initial signal request is made, the parameter initialization for the base nodes are performed and the antenna selection is performed. This is a random selection performed by comparison of base parameters, and it corresponds to the initial path taken by the ants. This stage of the call initializes the process of MCDM ACO. In general, the Ant Colony Optimization algorithm considers two parameters for performing the optimization process. The pheromone concentration τ and the evaporation rate ρ.

According to Dorigo's definition, the decision for the next edge to travel is performed in a probabilistic manner, using the relation:

Here,

$P_{ij}^k(t)$ = Probability that an ant k will travel from I to j node in graph at time t.

This probability is dependent on several factors like,

$\tau_{ij}(t)$ = Pheromone intensity at time t while travelling from node i to node j.

$\eta_{ij}(t)$ = Pheromone visibility at time t while travelling from node i to node j.

α = Importance of pheromone intensity.

β = Importance of pheromone visibility

$J_k(i)$ = Neighborhood set of node I for ant k.

Pheromone updates are dependent on relation,

$$P_{ij}^k(t) = \begin{cases} \frac{[\tau_{ij}(t)]^\alpha [\eta_{ij}]^\beta}{\sum_{l \in J_k(i)} [\tau_{ij}(t)]^\alpha [\eta_{ij}]^\beta}, & \text{if } j \in J_k(i) \\ 0, & j \notin J_k(i) \end{cases}$$
(2)

$$\Delta \tau_{ij}^k = \begin{cases} \frac{Q}{L_k}, & \text{if } (i,j) \epsilon \text{ tour} \\ 0, & \text{otherwise} \end{cases}$$
(3)

Here

$\Delta \tau_{ij}^k$ = Amount of pheromone updated by ant k while traveling from node i to j.

Q = Constant.

$L_k$ = Total path length covered by ant k.

Pheromone updates finally happens with this equation,

$$\tau_{ij}^k = \tau_{ij}^k + \Delta \tau_{ij}^k$$
(4)

and pheromone decay is dependent on the equation,

$$\tau_{ij} = (1 - \rho)$$
(5)

Here,

$\tau_{ij}$ = Pheromone intensity on edge joining node i and node j

ρ = Evaporation parameter of the graph.

The Multi Criteria Decision enabled Ant Colony Optimization process is used for decision making using multiple constraints rather than the single constraint of evaporation rate used on the pheromone availability. In our proposed method, two different levels of properties are considered before making a handoff. The first level of criteria deals with the current signal properties that the user is currently connected to and the second level of properties belong to the signals being offered for handoff.

The initial level properties (call based criteria) that are used are, type of traffic $\lambda_{tt}$ (throughput sensitive or delay sensitive), speed of travel $\lambda_s$, direction of travel $\lambda_d$, time to drop $\lambda_{td}$, handoff count $\lambda_{hc}$, packet loss rate $\lambda_{pl}$, latency $\lambda_l$, throughput $\lambda_{th}$, packet drop probability $\lambda_{pd}$, out of order delivery $\lambda_{od}$ and information priority $\lambda_{prio}$. Some of these properties correspond to the user's status, while others relate to the properties of the current being connected to the user. The next level of properties (signal based criteria) include cost of transmission $\lambda_c$, bandwidth $\lambda_{bw}$ and availability $\lambda_{avl}$.

Further, in Dorigo's method, the evaporation rate $\rho$ is a value common to the entire system. The modification in our proposal is brought about by using different evaporation rates for each of the channels. The evaporation rates depend on the parameters discussed above. If a channel is not available, then its visibility $\beta$ drops to 0, hence the current research is focused on the evaporation rate. A theoretical proof that proves the accuracy of the defined method is discussed below.

**Statement:** *Given a graph G(V,E), where V is the set of vertices that corresponds to the signal base and the first vertex $v_0$ corresponds to the user, and E is the edges or connections defining connectivity between the base and user. $\tau_{ij}$ represents the pheromone concentration, $\rho_{ij}$ represents the evaporation parameter, $\lambda_{tt}$ represents the type of traffic, $\lambda_s$ represents the speed, $\lambda_{td}$ represents the time to drop, $\lambda_{hc}$ represents the handoff count, $\lambda_{pl}$ represents the packet loss rate, $\lambda_l$ represents the latency, $\lambda_{th}$ represents the throughput, $\lambda_{pd}$ represents the packet drop probability, $\lambda_{prio}$ represents the priority of the current transmission, $v_c$ represents the cost of transmission and $v_{bw}$ represents the bandwidth of the channels, then the probability of selection of the edge ij by an ant is directly proportional to $\lambda_{tt}, \lambda_s, \lambda_{prio}, \lambda_{hc}, v_{bw}, \lambda_{th}$ and $\rho$, and is inversely proportional to $\lambda_{td}, \lambda_l, \lambda_{pd}, \lambda_{pl}$ and $v_c$.*

**Proof:**

According to equation (2),

$$P_{ij} \alpha \tau_{ij} \tag{6}$$

The probability of an edge to be selected is directly proportional to the pheromone deposits $\tau_{ij}$ in that edge.

The pheromone deposit rate is determined by two factors. The first being the number of ants traversed that path, the other being the evaporation rate $\rho_{ij}$.

$$\tau_{ij} \alpha \frac{1}{\rho_{ij}} \tag{7}$$

By default the evaporation rate is defined in the initial stage of the algorithm and does not change, while in this proposal, we provide different evaporation rates corresponding to the base stations, hence i and j represents the source and destinations respectively. Hence we can build the following equations

$$\tau_{ij} \alpha \lambda_{tt} \tag{8}$$

$$\tau_{ij} \alpha \lambda_s \tag{9}$$

$$\tau_{ij} \alpha \frac{1}{\lambda_{td}} \tag{10}$$

$$\tau_{ij} \alpha \lambda_{hc} \tag{11}$$

$$\tau_{ij} \alpha \frac{1}{\lambda_{pl}} \tag{12}$$

$$\tau_{ij} \alpha \frac{1}{\lambda_l} \tag{13}$$

$$\tau_{ij} \lambda_{th} \tag{14}$$

$$\tau_{ij} \alpha \frac{1}{\lambda_{pd}} \tag{15}$$

$$\tau_{ij} \alpha \lambda_{prio} \tag{16}$$

$$\tau_{ij} \alpha \frac{1}{v_c} \tag{17}$$

$$\tau_{ij} \alpha v_{bw} \tag{18}$$

Equations (8) to (18) show the relationship between the pheromone deposit and the additional defined properties. As the values $\lambda_{tt}, \lambda_s, \lambda_{hc}, \lambda_{th}, \lambda_{prio}$ and $v_{bw}$ tend to increase, the evaporation rate decreases, which implies that these properties are inversely proportional to the evaporation rate. Hence by equation (7),

$$\tau_{ij} \alpha \lambda_{tt} . \lambda_s . \lambda_{hc} . \lambda_{th} . \lambda_{prio} \tag{19}$$

As the values $\lambda_{td}, \lambda_l, \lambda_{pl}, \lambda_{pd}$ and $v_c$ tend to increase, the evaporation rate tends to increase, which implies that these properties are directly proportional to the evaporation rate. Hence by equation (7),

$$\tau_{ij} \alpha \frac{1}{\lambda_{td}.\lambda_l.\lambda_{pl}.\lambda_{pd}.v_c} \quad (20)$$

By combining equations (19) and (20),

$$\tau_{ij} \alpha \frac{\lambda_{tt}.\lambda_s.\lambda_{hc}.\lambda_{th}.\lambda_{prio}.v_{bw}}{\lambda_{td}.\lambda_l.\lambda_{pl}.\lambda_{pd}.v_c} \quad (21)$$

Hence, by combining equations (20) and (21), equation (6) can be written as

$$P_{ij} \alpha \frac{\lambda_{tt}.\lambda_s.\lambda_{hc}.\lambda_{th}.\lambda_{prio}.v_{bw}}{\lambda_{td}.\lambda_l.\lambda_{pl}.\lambda_{pd}.v_c} \quad (22)$$

From equation (22) we can imply the relationship between various properties used in the system with respect to $P_{ij}$ (Probabilty of selecting a route from i to j).

## V. RESULTS AND DISCUSSION

Performance analysis was performed using a single set of properties for the user, and various ant counts. The pheromone levels for CDMA, WiFi and WiMAX are measured for each of the ant values and tabulated. The values of ants range from 3 to 8. From the graphs (Figure: 1 to 7) we can see the convergence of the ants at different levels.

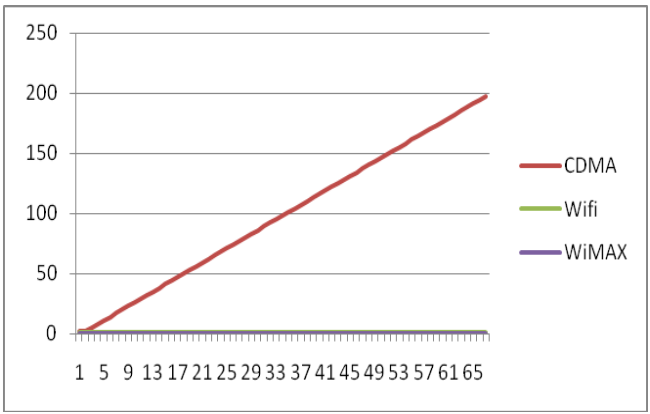

Fig. 1. Pheromone Levels with Ant Count 3

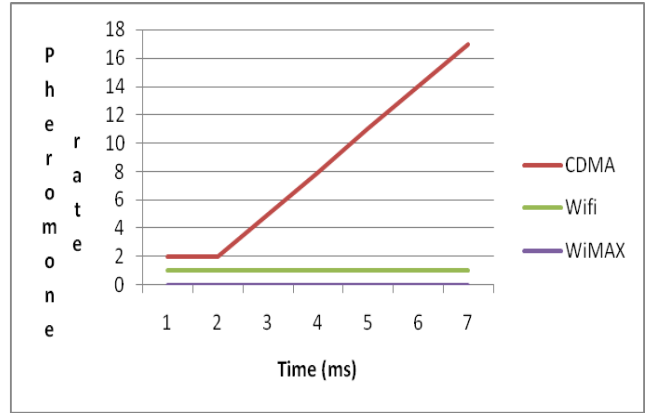

Fig. 2. Pheromone Levels with Ant Count 3 (A Closer Analysis)

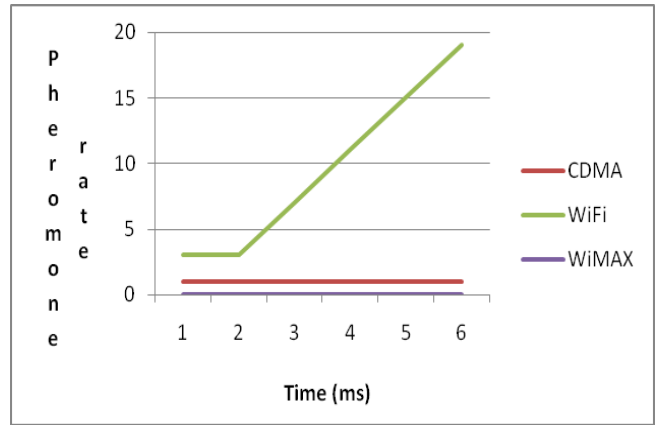

Fig. 3. Pheromone Levels with Ant Count 4 (A Closer Analysis)

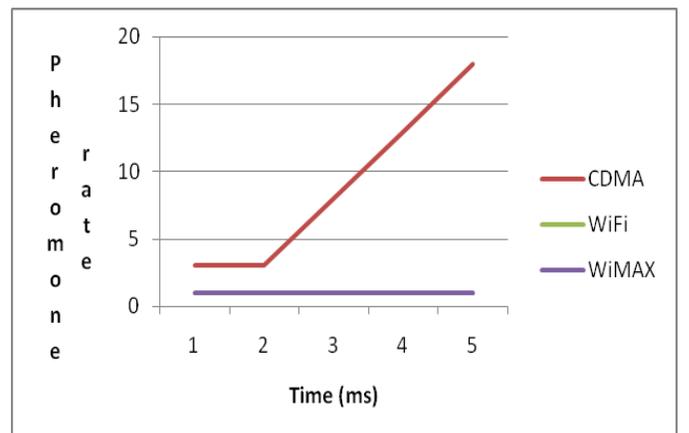

Fig. 4. Pheromone Levels with Ant Count 5 (A Closer Analysis)

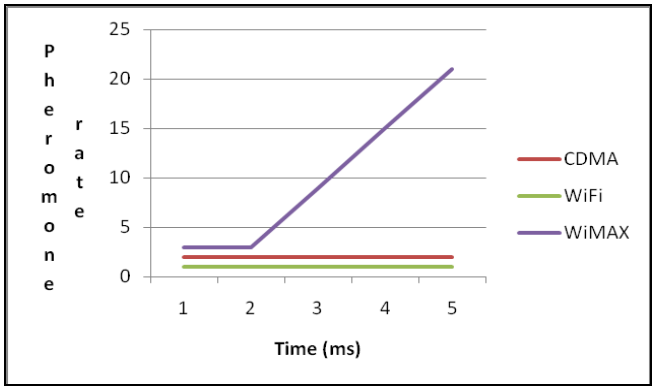

**Fig. 5. Pheromone Levels with Ant Count 6 (A Closer Analysis)**

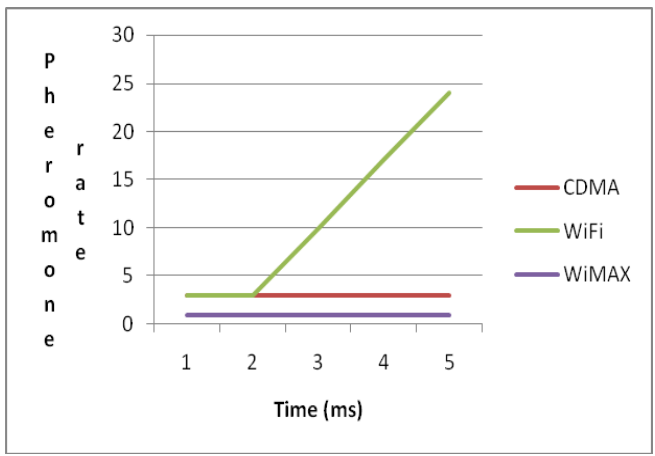

**Fig. 6. Pheromone Levels with Ant Count 7 (A Closer Analysis)**

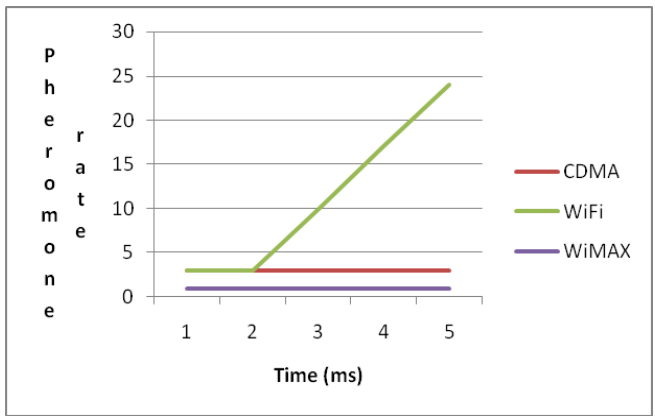

**Fig. 7. Pheromone Levels with Ant Count 8 (A Closer Analysis)**

From the graphs (Figure 1-7) we can say that the convergence is independent of the number of ants. But the time of conversion varies with respect to the number of ants.

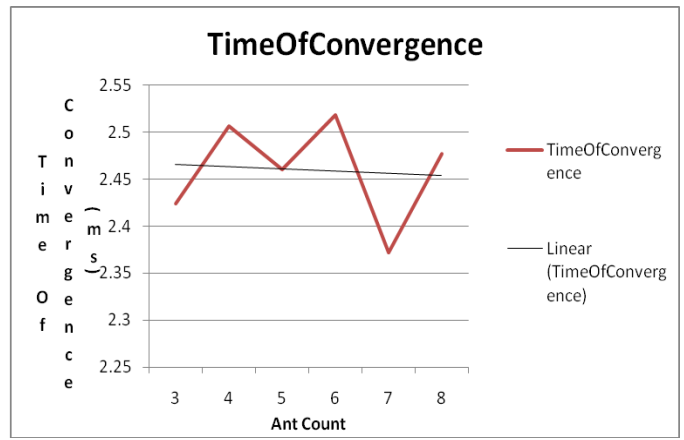

**Fig. 8. Time Of Convergence**

Figure 8 shows the time of convergence with respect to the ant count. The best level is recorded at an ant count of 7. The linear time of convergence shows that at any ant count, an average of 2.45 ms is maintained. Hence we can say that our proposed system converges efficiently with a very minimum convergence rate.

## VI. CONCLUSION

Due to the multi constrained nature of the handoff decision mechanism, a multi criteria decision making mechanism suits our current process. Further, the final handoff point determination is not expected to be accurate, but it is heavily time constrained. Hence the Ant Colony Algorithm with Multi Criteria Decision Making process works best for fast and efficient signal selection. Incorporation of Fuzzy Logic and Fuzzy based Decision making can be performed in future for result improvisation.

This method can be further optimized by using a combination of algorithms for the decision making. Due to the growth of parallelization technologies, the handoff decision making process can be parallelized, hence this will lead to faster solutions. Due to the intrinsic parallelizable nature of the ACO, this is made possible. This can be further improved by using cross validation based approaches for multi-core systems. i.e. Performing the process using two different methods and finally validating the results by cross referencing or aggregation. Consensus based Group Decision Making can be used to integrate the solutions. The group decision making can be performed in two ways, by using AHP by varying the importance of parameters or by using the Quantum approach. Using the quantum approach, a wavelet is constructed and the interference is used to find the aggregate result that can be used for the decision making process.